\documentclass[twocolumn,showpacs,preprintnumbers,amsmath,amssymb]{revtex4}

\usepackage{graphicx}
\usepackage{dcolumn}
\usepackage{bm}

\begin{document}

\preprint{APS/123-QED}

\title{The role of bulk energy in nuclear multifragmentation.}

\author{N.~Buyukcizmeci$^{1}$, A.S.~Botvina$^{2,3}$, I.N.~Mishustin$^{3,4}$
and R.~Ogul$^{1}$}

\affiliation{$^1$Department of Physics, University of Sel\c{c}uk,
42079 Konya, Turkey} \affiliation{$^2$Institute for Nuclear
Research, Russian Academy of Sciences, 117312 Moscow, Russia\\
$^3$Frankfurt Institute for Advanced Studies, J.W. Goethe
University, D-60438 Frankfurt am Main, Germany}
\affiliation{$^4$Kurchatov Institute, Russian Research Center,
123182 Moscow, Russia}

\date{\today}

\begin{abstract}
Because of thermal expansion and residual interactions, hot
nuclear fragments produced in multifragmentation reactions may
have lower nucleon density than the equilibrium density of cold
nuclei. In terms of liquid-drop model this effect can be taken
into account by reducing the bulk energy of fragments. We study
the influence of this change on fragment yields and isotope
distributions within the framework of the statistical
multifragmentation model. Similarities and differences with
previously discussed modifications of symmetry and surface
energies of nuclei are analyzed.
\end{abstract}


\maketitle

\section{Introduction}
\vspace{0.5cm}

Multifragmentation has been observed in nearly all types of high
energy nuclear interactions induced by hadrons, photons, and heavy
ions (see a review \cite{SMM}). This is an universal phenomenon
occurring when a large amount of energy is deposited in a nucleus,
and a hot blob of nuclear matter is formed. At low excitation
energies the nuclear system can be treated as a compound nucleus
\cite{Bohr}, which decays later via evaporation of light particles
or fission. However, at high excitation energy, or possibly,
compression during the initial dynamical stage of the reaction,
this matter blob will expand to the sub-saturation densities,
where it becomes unstable and breaks up into many fragments. As is
well known (see e.g. Refs. \cite{Viola,Karna}) multifragmentation
is a fast process, with a characteristic time around 100 fm/c.
Nevertheless, as shown by numerous analysis of experimental data,
a high degree of equilibration can be reached in these reactions,
and statistical models are very suitable for description of
fragment yields \cite{Botvina90,ALADIN,EOS,MSU,INDRA,FASA,Dag}. We
believe that taking multifragmentation into account is crucially
important for correct description of fragment production in high
energy reactions. On the other hand multifragmentation opens a
unique possibility for investigating the phase diagram of nuclear
matter at temperatures $T \approx 3-8$ MeV and densities around
$\rho \approx 0.1-0.3 \rho_0$ ($\rho_0 \approx 0.15$ fm$^{-3}$ is
the normal nuclear density). These conditions are typical for the
liquid-gas coexistence region. It is interesting that similar
conditions are realized in stellar matter during the supernova
explosions \cite{Botvina04,Botvina05}.

In the course of nuclear disintegration hot primary fragments are
first formed in close vicinity to each other, and, therefore, they
are still subject to Coulomb and, possibly, residual nuclear
interactions. It is commonly accepted that the liquid-drop
description of individual nuclei is very successful in nuclear
physics. However, in a multi-fragment system in the freeze-out
volume the parameters of the liquid-drop model may change as
compared with those for isolated nuclei. An obvious example is the
reduction of the fragment Coulomb energy due to the presence of
other fragments. This effect can be reasonably evaluated within
the Wigner-Seitz approximation \cite{SMM}. The Coulomb interaction
between the fragments can also influence the proton and neutron
distributions in hot fragments \cite{Jandel}. Moreover, nuclear
interactions parameterized as bulk, surface and symmetry energy
terms in the liquid-drop description of nuclei may change too.
Possible modifications of surface and symmetry energies of primary
fragments, and constraints from relevant experimental data, were
analyzed in the previous works
\cite{Botvina06,nihal,LeFevre,Iglio,Souliotis}. It was found that
the symmetry energy of hot fragments drops significantly in the
freeze-out volume, and the surface energy can be considerably
modified at high temperatures. In this paper we investigate
possible changes of the nuclear bulk energy in primary fragments,
and how this may affect their production in multifragmentation
reactions. We will show that these effects may be quite essential
for explaining some key observables, and they should be included
in realistic models.

\vspace{0.5cm}

{\bf 2. Statistical description of nuclear multifragmentation}

\vspace{0.5cm}

All dynamical models used for description of the initial stage of
the reaction lead to the conclusion that after a time interval of
few tens fm/c, when fast particles leave the system, the evolution
of the remaining nuclear system changes its character. Because of
intensive interactions between nucleons the system evolves toward
statistical equilibrium. At later times the hot nuclear residue
expands and breaks-up into hot primary fragments. The Statistical
Multifragmentation Model (SMM) is based on the assumption of
statistical equilibrium between produced fragments in a
low-density freeze-out volume \cite{SMM}. We believe that at this
point the chemical equilibrium is established, i.e., the baryon
composition (mass and charge) of primary fragments is fixed.
However, the fragments can still interact with other nuclear
species via the Coulomb and nuclear mean fields. Hence their
energies and densities may be affected by these residual
interactions. All breakup channels composed of nucleons and
excited fragments are considered, and the conservation of mass,
charge, momentum and energy is taken into account. An advantage of
the model is that the formation of a compound nucleus is included
as one of the channels. This allows for a smooth transition from
the decay via evaporation and fission at low excitation energies
\cite{Bohr} to the multifragmentation at high excitations. In the
microcanonical treatment \cite{SMM,Botvina01} the statistical
weight of the decay channel $j$ is given by $W_{\rm j} \propto
exp~S_{\rm j}$, where $S_{\rm j}$ is the entropy of the system in
channel $j$ which is a function of the excitation energy $E_{\rm
x}$, mass number $A_{0}$, charge $Z_{0}$ and other global
parameters of the source. After formation in the freeze-out
volume, the fragments propagate independently in their mutual
Coulomb field and undergo secondary decays. The deexcitation of
the hot primary fragments proceeds via evaporation, fission, or
Fermi-breakup \cite{Botvina87}.

In the SMM light fragments with mass number $A\le 4$ and charge
$Z\le 2$ are considered as structure-less particles (nuclear gas)
with masses and spins taken from the nuclear tables. Only
translational degrees of freedom of these particles contribute to
the entropy of the system. Fragments with $A > 4$ are treated as
heated nuclear liquid  drops, and their individual free energies
$F_{AZ}$ are parameterized as a sum of the bulk, surface, Coulomb
and symmetry energy terms:
\begin{equation}
F_{AZ}=F^{B}_{AZ}+F^{S}_{AZ}+E^{C}_{AZ}+E^{sym}_{AZ}.
\end{equation}

In this standard expression $F^{B}_{AZ}=(-W_0-T^2/\epsilon_0)A$ is
the bulk energy term including contribution of internal
excitations controlled by the level-density parameter
$\epsilon_0$, and $W_0 = 16$~MeV is the binding energy of infinite
nuclear matter.
$F^{S}_{AZ}=B_0A^{2/3}((T^2_c-T^2)/(T^2_c+T^2))^{5/4}$ is the
surface energy term, where $B_0=18$~MeV is the surface coefficient
at $T=0$, and $T_c=18$~MeV is the critical temperature of infinite
nuclear matter. The Coulomb energy is $E^{C}_{AZ}=cZ^2/A^{1/3}$,
where $c$ is the Coulomb parameter obtained in the Wigner-Seitz
approximation, $c=(3/5)(e^2/r_0)(1-(\rho/\rho_0)^{1/3})$, where
$e$ is the proton charge, $r_0$=1.17 fm, and the last factor
describes the screening effect due to presence of other fragments.
$E^{sym}_{AZ}=\gamma (A-2Z)^2/A$ is the symmetry energy term,
where $\gamma = 25$~MeV is the symmetry energy coefficient. These
parameters are taken from Bethe-Weizs\"acker formula and
correspond to the isolated fragments with normal nuclear density.
This assumption has been proven to be quite successful in many
applications. However, a realistic treatment of primary fragments
in the freeze-out volume may require certain modifications of the
liquid-drop parameters as indicated by experimental data.

In the grand canonical treatment of the SMM \cite{Botvina85},
after integrating out translational degrees of freedom, one can
write the mean multiplicity of nuclear fragments with $A$ and $Z$
as
\begin{eqnarray}
\label{naz} \langle N_{AZ} \rangle =
g_{AZ}\frac{V_{f}}{\lambda_T^3}A^{3/2} {\rm
exp}\left[-\frac{1}{T}\left(F_{AZ}(T,\rho)-\mu A-\nu
Z\right)\right]. \nonumber
\end{eqnarray}
Here $g_{AZ}$ is the ground-state degeneracy factor of species
$(A,Z)$, $\lambda_T=\left(2\pi\hbar^2/m_NT\right)^{1/2}$ is the
nucleon thermal wavelength, and $m_N \approx 939$ MeV is the
average nucleon mass. $V_f$ is the free volume available for the
translational motion of fragments. The chemical potentials $\mu$
and $\nu$ are found from the mass and charge constraints:
\begin{equation} \label{eq:ma2}
\sum_{(A,Z)}\langle N_{AZ}\rangle A=A_{0},~~ \sum_{(A,Z)}\langle
N_{AZ}\rangle Z=Z_{0}.
\end{equation}

As was demonstrated by numerous comparisons of the SMM with
various experiments, the model describes data very well (see,
e.g., Refs. \cite{Botvina90,ALADIN,EOS,MSU,INDRA,FASA,Dag}). This
confirms that the statistical approach with liquid-drop
description of individual fragments provides adequate treatment of
the multifragmentation process. This also justifies the
application of the statistical approach for investigating the
liquid-gas phase transition in nuclear systems
\cite{DasGupta,Bugaev}.

\vspace{0.5cm}

{\bf 3. Modifications of properties of primary fragments}

\vspace {0.5cm}

In recent years several new analyzes of experimental data with
statistical models, related to the nuclear isospin in
multifragmentation reactions have been performed
\cite{Botvina06,LeFevre,Iglio,Souliotis}. They conclude that
modifications of the liquid-drop parameters of hot fragments
produced in the freeze-out volume are needed to explain the data.
It was suggested that this can happen because of a new physical
environment where fragments are formed, in particular, since they
are surrounded by nucleons and other hot fragments. The residual
interactions may lead to energy and density changes which can
effectively be explained by a modification of the macroscopic
nuclear parameters. The symmetry energy coefficient $\gamma$ was
investigated in several independent experiments
\cite{LeFevre,Iglio,Souliotis}, which used both the isoscaling
phenomenon \cite{traut} and isotope distributions of fragments. It
is important that all experiments come to the conclusion that the
coefficient $\gamma$ drops from about 25 MeV, known for isolated
cold nuclei, down to $\approx 15$ MeV for hot primary fragments at
multifragmentation. The same results were extracted also from
analysis of mean neutron content of fragments
\cite{Botvina05,Souliotis}. One of the aims of future experiments
is to verify this conclusion.

A recent analysis of the ALADIN data has revealed modifications in
the nuclear surface properties too \cite{Botvina06}. There the
neutron-to-proton ($N/Z$) dependence of the surface energy was
analyzed for different event classes corresponding to different
excitation energies. At low excitation energies, corresponding to
the onset of multifragmentation, the surface energy follows the
trend predicted by the standard liquid-drop model, i.e., it
decreases with $N/Z$. This trend is usually explained by the
contribution of the surface part of the symmetry energy. In the
region of developed multifragmentation (temperatures
$T\approx$5--6 MeV), where intermediate-mass fragments (IMF:
$Z=3-20$) are mostly produced, the surface energy coefficient
becomes nearly independent of the $N/Z$ ratio. Taking into account
this result we conclude that subdivision of the total symmetry
energy into volume and surface parts becomes irrelevant at
multifragmentation conditions. We point out also that similar
conclusions about changing surface properties of nuclei were
obtained within the dynamical AMD model \cite{Ono04}. Therefore,
the observed decrease of the symmetry energy of fragments should
not be related to the increase of the total surface of fragments
at multifragmentation. It is more likely that this is an
indication of the medium modification of $\gamma$ coefficient in
the system of many fragments.

We should point out that there are attempts to explain the isospin
related observables within the mean-field models, as a result of
reduced mean density of the excited nuclear system \cite{Li06,
Xu07}. As is well known, the corresponding equations of state
predict decreasing symmetry energy at subnuclear densities.
However, within this approach the fragment formation is not
considered, and it is assumed that the produced fragments keep a
"memory" about symmetry energy of a low-density system. In our
opinion this assumption is not very convincing, since the fragment
formation process (e.g., associated with the liquid-gas phase
transition) may drastically change properties of the system, and
especially properties of primary fragments. We believe that it is
more reasonable to assume that primary fragments produced in an
expanded freeze-out volume have a reduced density, and this can
lead to reduction of their symmetry energy observed in the
experiments \cite{LeFevre,Iglio,Souliotis}.

Until now the nucleon density of individual fragments was not
explicitly considered in the SMM calculations, and, on default, it
was assumed that all fragments have normal nuclear density
$\rho_0$. However, as a result of residual interaction between the
fragments and their thermal expansion the density of fragments
($\rho_f$) may in fact become smaller than $\rho_0$. This will
lead to a reduction of the bulk binding energy $W_0$. For
estimation of this effect we use the following expression for the
bulk energy as a function of the nucleon density:
\begin{equation}
\label{eq:wrho}\frac{E_B}{A}(\rho_{f}) \equiv\ -W_0(\rho_{f})=-W_0
+ \frac{K}{18}\frac{(\rho_{f}-\rho_0)^2}{\rho_0^2}~,
\end{equation}
where $K \approx 260$ MeV is the nuclear compressibility modulus.
There are natural limits for possible reduction of the average
density of fragments. Due to the spinodal instabilities and
Coulomb fluctuations at the subnuclear densities $\rho < (0.6-0.7)
\rho_0$, uniform nuclear matter becomes unstable, and the nuclear
``pasta'' phases are produced \cite{Pethick,Brown,Lamb}. By this
reason, we take $\rho_{f} \approx 0.6\rho_0$ as the minimum
possible nuclear density in individual fragments. One can see from
eq.~(\ref{eq:wrho}), that this corresponds to decreasing $W_0$
from 16 MeV at $\rho_{f}=\rho_0$ down to $W_0 \approx 14$ MeV. In
the following analysis we allow for even smaller $W_0$, in order
to perform a complete investigation of this effect.

At the last stage of the multifragmentation process hot primary
fragments undergo deexcitation and propagate in the mutual Coulomb
field. As was demonstrated in many works (see, e.g., \cite{SMM})
this stage is very important for correct calculations of final
yields of fragments. In the beginning of the deexcitation the hot
fragments are still surrounded by other species, and, therefore,
their modified properties should be taken into account. As far as
we know, only one evaporation code was designed, which takes into
account the modified properties of fragments in their
de-excitation. It was developed in Refs. \cite{nihal,daniela},
where modifications of symmetry energy were explicitly considered.
In the present analysis we use the same prescription. Namely, we
start from the modified bulk energies of hot nuclei and restore
their normal properties by the end of the evaporation cascade. In
actual calculations we have used a simple interpolation between
these two limiting states. The energy and momentum conservation
laws were fulfilled in the course of this evaporation process.

We emphasize that the treatment of this de-excitation stage should
be consistent with the physical properties of primary fragments.
For example, a failure to describe the experimental isoscaling
data \cite{Liu04}, by using a dynamical model for primary fragment
formation, may be related to the fact that the sequential
evaporation from primary fragments was included without paying
attention to their reduced densities and, consequently, to their
lower symmetry energies. As was demonstrated in Refs. \cite{nihal,
Iglio}, this effect can influence essentially the observed isotope
distributions, since separation energies of neutrons and protons
are changed when compared to the isolated low-excited nuclei. The
transition to the secondary de-excitation can be consistently
controlled in statistical description of fragment formation, by
using a generalized evaporation prescription \cite{nihal}.
Moreover, we think that the final conclusions about properties of
primary fragments can be obtained only after a many-component
analysis of experimental data. Besides the isotope information and
isoscaling observable this should include the corresponding
fragment charge distributions, IMF multiplicities, temperatures,
and the other relevant characteristics (examples of such an
approach are presented in Refs. \cite{ALADIN, EOS, MSU, Botvina06,
Xi97}).

\vspace{0.5cm}

{\bf 4. Influence of the bulk energy on multifragmentation
characteristics}

\vspace{0.5cm}

We have used the SMM to simulate multifragmentation of the gold
source with excitation energies in the range of 2--12 MeV per
nucleon at different values of the bulk energy coefficient $W_0$.
For simplicity, we fix the freeze-out density at $1/3\rho_0$ as
was done also in many previous SMM studies. In Figure 1 we show
some characteristics of the hot fragments: effective temperatures,
mass number of the largest fragment, average multiplicity of IMFs
produced in the freeze-out volume. The effective temperature
$T_{eff}$ is calculated from the energy balance in the system, in
the same way as it was done in Ref. \cite{Botvina06}.

\begin{figure} [tbh]
\begin{center}

\includegraphics[width=8cm,height=12cm]{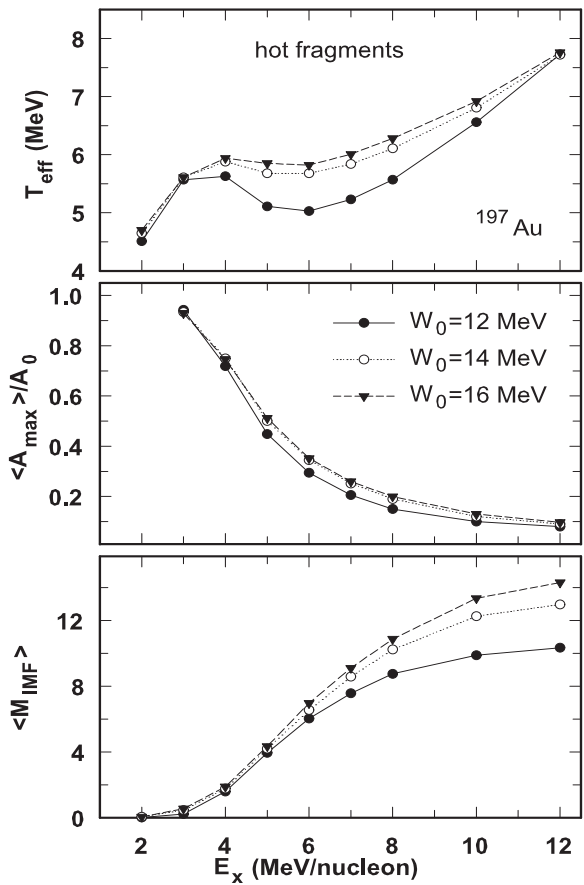}

\caption{\small{Effective temperature, reduced mass number of the
maximum fragments and average multiplicity of the intermediate
mass fragments as a function of the excitation energy at different
bulk energy coefficients $W_0=12, 14, 16$~MeV are shown,
respectively, in the top, middle and the bottom panels.}}

\end{center}
\end{figure} 

\begin{figure} [tbh]
\begin{center}

\includegraphics[width=8cm,height=12cm]{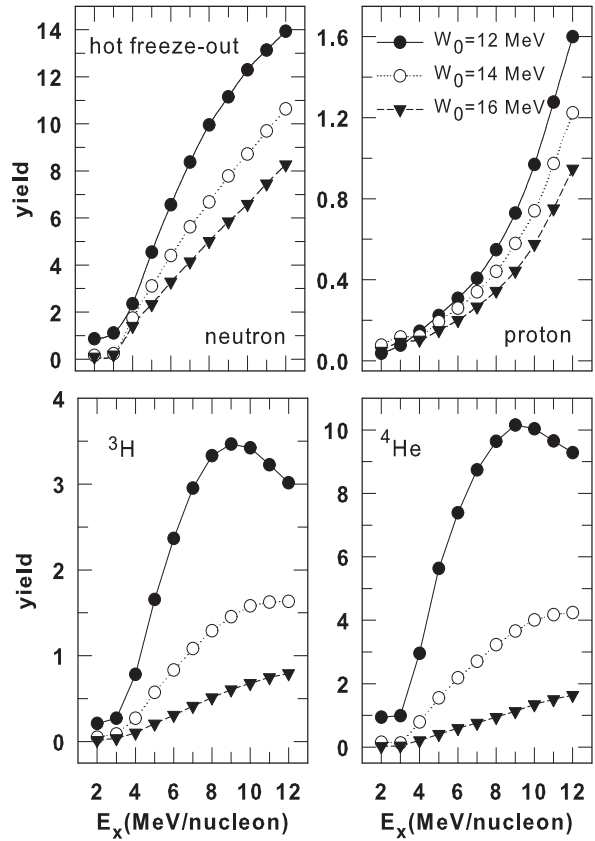}

\caption{\small{Yields of the neutrons, protons, tritons  and
alpha particles in the hot freeze-out volume as a function of the
excitation energy at different bulk energy coefficients. }}

\end{center}
\end{figure} 

\begin{figure} [tbh]
\begin{center}

\includegraphics[width=8cm,height=12cm]{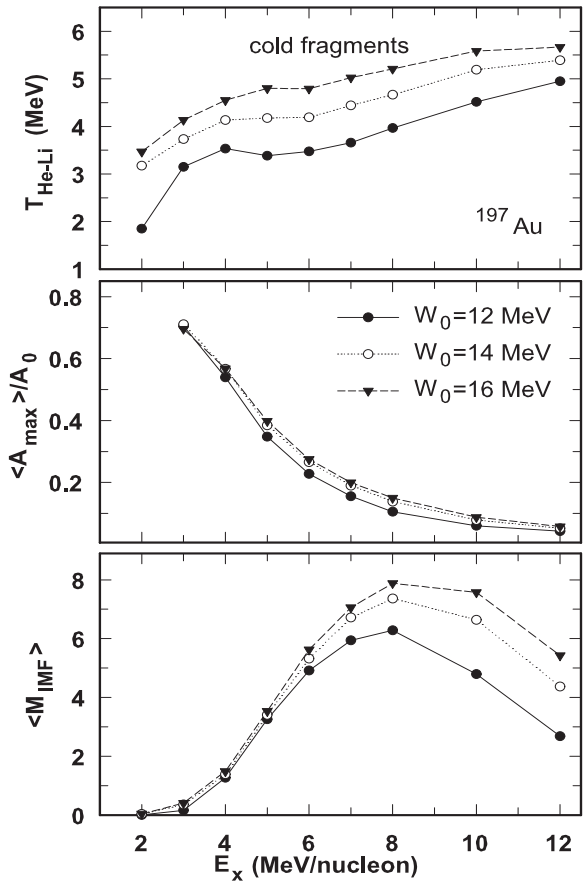}

\caption{\small{Characteristics of the final cold fragments (after
secondary deexcitation) produced from Au sources. The notations
are the same as in Fig.~1, here in the top panel variations of the
isotopic Helium-Lithium temperatures $T_{He-Li}$ are also shown.}}

\end{center}
\end{figure} 

\begin{figure} [tbh]
\begin{center}

\includegraphics[width=8cm,height=12cm]{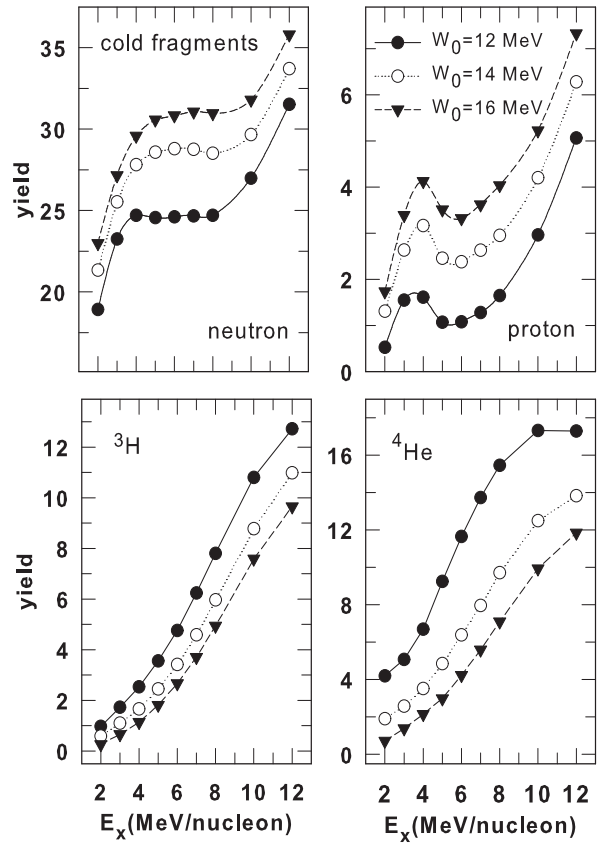}

\caption{\small{Yields of the neutrons, protons, tritons  and
alpha particles for final cold fragments, after secondary
deexcitation.}}

\end{center}
\end{figure} 

\begin{figure} [tbh]
\begin{center}

\includegraphics[width=8cm,height=12cm]{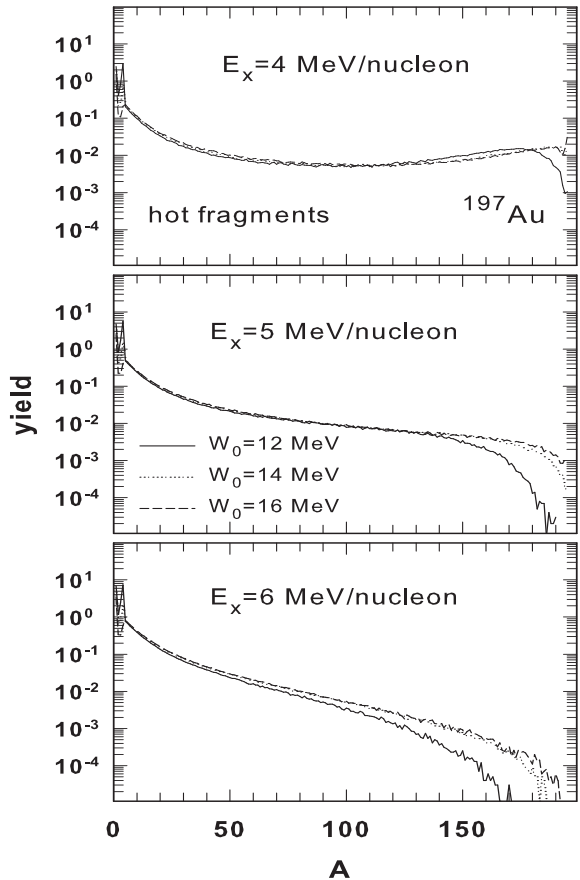}

\caption{\small{Mass distributions of hot fragments produced from
Au sources at excitation energies of 4, 5 and 6 MeV per nucleon
for different bulk energy coefficients.}}

\end{center}
\end{figure} 

\begin{figure} [tbh]
\begin{center}

\includegraphics[width=8cm,height=12cm]{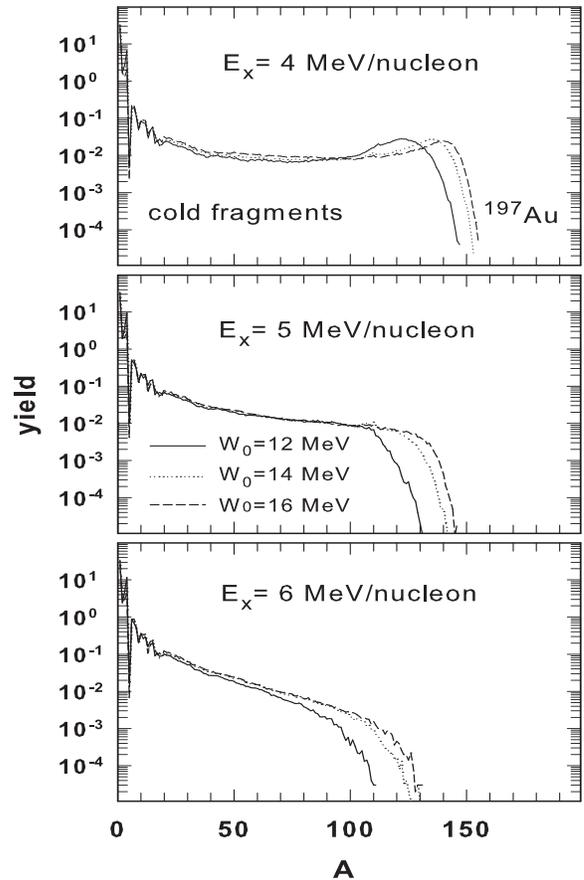}

\caption{\small{Mass distributions of the cold fragments produced
from Au sources at excitation energies of 4, 5 and 6 MeV per
nucleon.}}

\end{center}
\end{figure} 

\begin{figure} [tbh]
\begin{center}

\includegraphics[width=8cm,height=12cm]{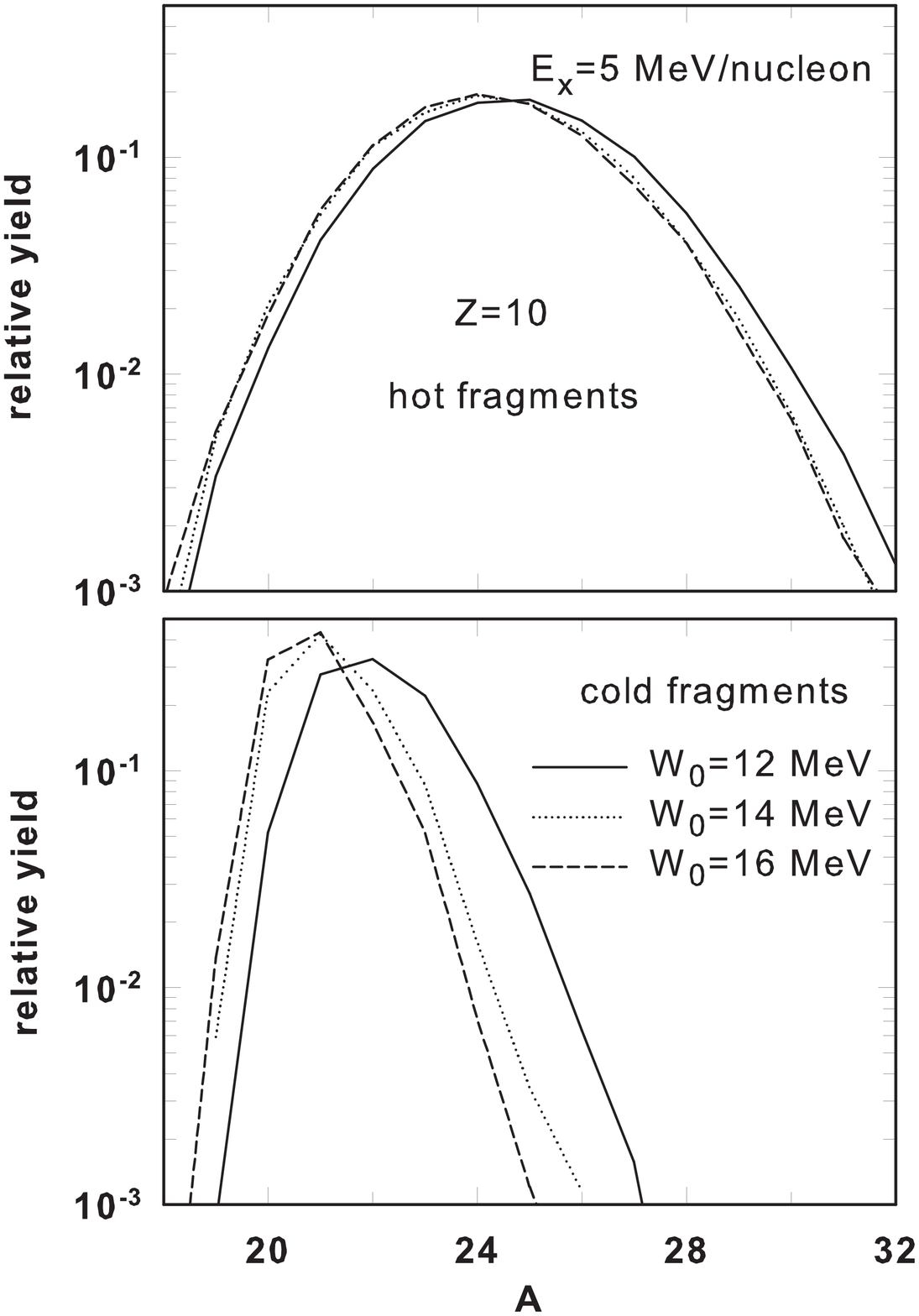}

\caption{\small{Mass distributions of hot and cold fragments with
the charge number Z=10 at the excitation energy of 5 MeV per
nucleon.}}

\end{center}
\end{figure} 

One can see that decreasing the bulk energy does not change
qualitatively the general picture of multifragmentation.
Nevertheless, some new important features appear. For example, the
back-bending of the caloric curve become more pronounced, i.e.,
multifragmentation reactions become more endothermic. Such
behavior of $T_{eff}$ can be interpreted as a manifestation of the
negative heat capacity. An interesting result is that the number
of IMFs and the mass of large fragments at freeze-out become
smaller, while the number of light particles (Fig.~2) is increased
considerably. From Fig.~2 one can see a drastic rise of the yields
of $\alpha$-particles and clusters with $A < 4$ at reduced $W_0$.
The reason is that we take into account fragments with $A \leq 4$
as 'gas' particles with their table binding energies, while
binding energies of other fragments become smaller. We believe
that $\alpha$-particles have normal properties in the low density
freeze-out volume.  This assumption is consistent with many
experimental data. For example, by interpretation of fragment
yields in direct knock-out reactions on nuclei, $\alpha$- clusters
are assumed to pre-exist even at the normal nuclear density, which
is much higher than the freeze-out density at multifragmentation
\cite{Boal}. We suggest that this effect may explain an excess of
$\alpha$- particle yields over SMM predictions observed in
emulsion data \cite{Tashkent}.

It is important that all new trends in the fragment yields caused
by changing the bulk energy survive after the secondary
deexcitation. Figs. 3 and 4 show the final yields of cold
fragments. The number of emitted $\alpha$- particles even
increases, since their evaporation becomes more probable at the
first stages of deexcitation, in addition to the increased
production at the break-up of the source (see Fig.~4). This fact
leads to the decrease of the so-called He--Li temperature (see
Fig.~3), which is one of the useful experimental observable for
the calorimetry of multifragmentation processes
\cite{Pochodzalla,Bondorf98}. On the other hand, decreasing $W_0$
causes a drop of neutron and proton yields, since they are now
bound in the light clusters. For the same reason, the maximum IMF
multiplicity becomes lower.

In order to give a detailed picture of nuclear break-up, in Figs.
5 and 6 we present the distributions of produced fragments in the
full mass range, before and after the secondary deexcitation. We
choose the range of excitation energies where the transition from
the 'U-shaped' to the power-law distributions is taking place. It
is obvious that decreasing the bulk energy favours
multifragmentation, since less bound largest clusters are
destroyed in favor of light clusters and nucleons. However, the
yields of IMFs in this crucial region change very little with
$W_0$ (as one can also see from Figs. 1 and 3), and the general
shape of the IMF distributions does not change. By using a
well-known power-law parametrization $A^{-\tau}$ of IMF yields we
obtain that the $\tau$ remains practically the same for the
considered variations of $W_0$. This result is quite
understandable: since the $W_0$ parameter enters binding energies
of all IMF, their relative yields do not change. For this reason,
the previously reported results concerning evolution of the
isospin-dependent contribution to the surface energy with
excitation energy of the system \cite{Botvina06}, which are based
on consideration of the IMF yields, remain valid in the case of
reduced bulk energy too.

Much attention is now paid to the isotope production in
multifragmentation reactions, because of its connection to
properties of neutron-rich nuclear matter and to astrophysical
applications \cite{Botvina05}. We have investigated this problem,
and Fig.~7 demonstrates isotope distributions of fragments with
$Z=10$ before (in the freeze-out volume) and after the secondary
deexcitation (at infinity). At smaller bulk energies the hot
fragments become a little bit more neutron rich. This is explained
by enhanced production of the symmetric $\alpha$ clusters, leading
to a neutron enrichment of the remaining nuclear matter. However,
this trend is more evident for the cold fragments, and we have two
contributions to this effect. The first one is caused by the
considerable probability of $\alpha$ particles emission in the
beginning of the evaporation cascade, due to the modified bulk
energy of heavier nuclei. The second one is related to the fact
that the initial temperature is lower, and, therefore, the
evaporational evolution of nuclei towards the $\beta$-stability
line is ceased at relatively large neutron excess. This
possibility for obtaining neutron-rich isotopes should be
considered in future analyzes of data alongside with possible
reduction of the symmetry energy of fragments
\cite{nihal,LeFevre,Iglio,Souliotis,traut}.

On the other hand, these two possibilities may complement each
other, since an expansion (i.e., a reduced density) of hot
fragments should lead to decreasing their symmetry energy too
\cite{ditoro}. The bulk symmetry energy as a function of density
is usually parameterized as
\begin{equation}
\label{eq:gamma} \gamma(\rho_{f}) \propto \gamma(\rho_0)
(\frac{\rho_{f}}{\rho_0})^{n}~,
\end{equation}
where the exponent $n$ ranges between 0.5 and 1.5 depending on the
model assumptions on isospin-dependent nuclear interactions.
Recent analyses of experimental data have shown that around
$\rho_{0}$ the values of $n$ may lie in the interval between 0.7
\cite{yennello} and 0.9 \cite{BALI}. Thus, in order to explain a
considerable drop of 40\% in $\gamma$, which fits experimental
observations, we need to decrease $\rho_{f}$ to $(0.45 - 0.55)
\rho_0$. As we have noted previously, this is already lower than
the threshold for appearance of different ``pasta'' phases
\cite{Pethick,Brown,Lamb}. Also, the realistic Hartree--Fock and
Thomas--Fermi calculations for hot isolated compound nuclei
predict a rather moderate expansion at temperatures of 5 -- 6 MeV,
not more than 10 -- 20\% \cite{Bonche,Suraud}, depending on the
nuclear forces used. Therefore, the maximum reduction of the
fragment density, which can be approximately considered, is around
30--40\%. In this situation one should find additional mechanisms
to explain the observed modification of the symmetry energy.

\vspace{0.5cm}

{\bf 5. Conclusions}

\vspace{0.5cm}

In this paper we proceed further with the investigation of
possible modifications of nuclear properties in surrounding of
other nuclei, which can still interact with nuclear and
electromagnetic forces. This subject is relevant for nuclear
matter at low density, which breaks-up into fragments. This state
of matter can also be considered as a mixed phase of the nuclear
liquid-gas phase transition. It is expected that this transition
takes place in such astrophysical processes as collapses of
massive stars and supernova explosions. In terrestrial
laboratories this physical phenomenon can experimentally be
studied in multifragmentation reactions induced by hadrons and
heavy ions of intermediate energies. Actually, the
multifragmentation is a unique tool, which extends our
conventional study of isolated compound nuclei to the nuclei
embedded in the environment of many nuclear species.

In recent years interesting new analyzes of multifragmentation
data have appeared which indicate a significant modification of
properties of nuclei in such a medium. They include: decrease of
the symmetry energy of nuclei, disappearance of an
isospin-dependent contribution to the surface energy, and possible
expansion of hot nuclei. In the present work we have investigated
the influence of decreasing bulk energy on the fragment
production. We have found that this effect should lead to an
enhanced production of $\alpha$ clusters. In the case of a
neutron-rich system, this should result in larger neutron richness
of heavy fragments. Another important finding is that, despite of
disintegration of the system into larger number of fragments, the
shape of the IMF mass distributions does not change with
decreasing bulk energy. Also we have found a moderate modification
of the caloric curve, namely a more pronounced back-bending, which
indicates a negative heat capacity, and a drop of the apparent
helium-lithium temperature. It is important that all findings are
consistent with previous results on modification of fragment
properties extracted from experimental data. Moreover, the
expansion of hot fragments may provide an additional contribution
to some effects observed earlier, such as decrease of their
symmetry energy, and increase of neutron richness of cold
fragments. We foresee new analyzes of experimental data in order
to clarify contributions of different mechanisms leading to the
modification of fragment properties in multi-particle environment.

This work was partly supported by DFG grant 436RUS 113/711/02
(Germany) and by grant NS-8756.2006.2 (Russia). A.S.B. thanks FIAS
(Frankfurt/Main) for hospitality and support. N.B. thanks
Sel\c{c}uk University-Scientific Research Projects (BAP) for
financial support.

\end{document}